\documentclass[twocolumn,showpacs,preprintnumbers,aps,prl,amsmath,amssymb]{revtex4}

\usepackage[english]{babel}
\usepackage[latin2]{inputenc}
\usepackage[T1]{fontenc}

\usepackage[usenames]{color}

\usepackage{ae,aecompl}
\usepackage{enumerate}
\usepackage{epsfig}
\usepackage{graphics}
\usepackage{graphicx}
\usepackage{amsmath}
\usepackage{amssymb}
\usepackage{pifont}
\usepackage[sort&compress]{natbib}
\usepackage{wasysym}
\usepackage{color}

\def\ev #1{\left\langle #1 \right\rangle}

\begin{document}

\title{Fluctuation scaling versus gap scaling}
\author{Zolt\'an Eisler}
\email{eisler@maxwell.phy.bme.hu} \homepage{http://maxwell.phy.bme.hu/~eisler}
\author{J\'anos Kert\'esz}
\altaffiliation[Also at ]{Laboratory of Computational Engineering, Helsinki University of Technology, Espoo, Finland} 
\affiliation{Department of Theoretical Physics, Budapest University of Technology and Economics, Budapest, Hungary}

\date{\today}
\pacs{87.23.--n, 89.75.-k, 89.75.Da, 05.40.-a}

\begin{abstract}
Fluctuation scaling is observed phenomenon from complex networks through finance to ecology. It means that the variance and the mean of a specific quantity are related as $\ev{\sigma^2|n}\propto \ev{n|A}^{2\alpha}$ with $1/2\geq \alpha \geq 1$ when a parameter $A$ (usually the system size) is varied. $A$ can be the strength of the node, the capitalization of the firm or the area of the habitat. On the other hand, quantities often obey gap scaling meaning that their density function depends on, say, the system size $A$ as $P(n|A) = n^{-1}F(n/A^{\Phi})$. This note describes that these two notions cannot coexist except when $\alpha = 1$. In this way one can empirically exclude the possibility of gap scaling in many complex systems including population dynamics, stock market fluctuations and Internet router traffic, where $\alpha < 1$.
\end{abstract}

\maketitle

Scaling has a fundamental importance in statistical physics. It has found countless successful applications starting with critical phenomena \cite{stanley.phase}, but more recently also outside the classical domain of physics, for example in ecology \cite{banavar.ecology}. The discovery of self-organized criticality \cite{bak.soc} has even shown that in certain systems these features are not tied to a special, critical set of parameters, but they reflect the generic behavior. Mono-scaling or gap scaling (in contrast to multi-scaling) means that the probability distribution of a quantity $n$ depends on the parameter $A$ as
\begin{equation}
	P(n|A) = n^{-1}F\left (\frac{n}{A^{\Phi}}\right),
	\label{eq:gapscaling}
\end{equation}
and this form can account for a number of observations about power law behavior in real systems.

\begin{figure}[!th]
\centerline{\includegraphics[width=230pt,trim=10 20 10 10]{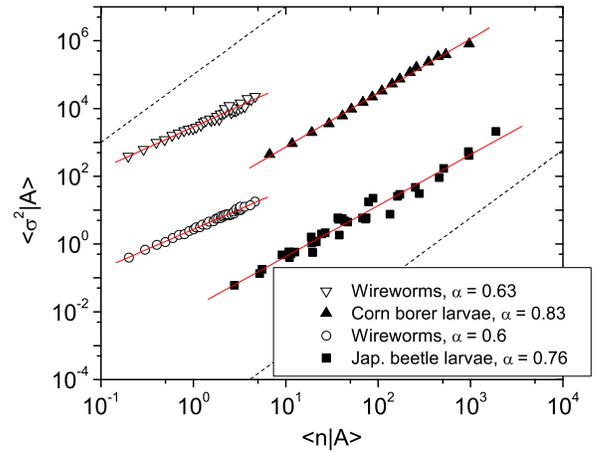}}
\caption{Variance versus mean of population abundance for four species (points were shifted for better visibility). A group presents the same species, but different area sizes. The form $\ev{\sigma^2|A} \propto \ev{n|A}^{2\alpha}$ was fitted, $\alpha$ values are given in the plot, typical errors are $\pm 0.03$. The dashed lines indicate $\alpha = 1$.}
\label{fig:fs}
\end{figure}

Another notion for complex systems originates from ecology where it is called Taylor's law, and it can be generally termed as \emph{fluctuation scaling}. In $1961$ Taylor \cite{taylor} showed that for many species, when varying the habitat area $A$ (a kind of system size) the variance $\ev{\sigma^2|A}$ and the mean $\ev{n|A}$ of the population level are related as
\begin{equation}
	\ev{\sigma^2|A} = \ev{n^2|A}-\ev{n|A}^2 \propto \ev{n|A}^{2\alpha}.
	\label{eq:taylor}
\end{equation}
This relationship constitutes one of the few quantitative laws with general validity in ecology. Since this work many observations showed that $\alpha$ is species specific and it is predominantly in the range $1/2-1$, see Fig. \ref{fig:fs} for examples \cite{taylor.insect, kilpatrick.ives}. It can be shown that $\alpha = 1/2$ and $\alpha = 1$ are limiting cases. The former can be the consequence of the central limit theorem, while the latter is attributed to a kind of synchronization \cite{anderson.variability}. 

Moreover, Eq. \eqref{eq:taylor} has been observed for a broad range of positive quantities ranging from Internet router traffic through web page visitations to the stock market \cite{barabasi.fluct, duch.internet, eisler.non-universality}. In these cases $A$ is -- instead of habitat area -- some other size-like parameter, such as the importance of the router/web page, or the capitalization of the stock. Often $\alpha = 1/2$ or $1$ were found but there exist numerous counterexamples as well: for the stock market \cite{eisler.non-universality} and for the Internet traffic \cite{duch.internet} intermediate exponents were observed. 

Both gap scaling and fluctuation scaling characterize a large number of complex systems. Nevertheless, for the same quantity \emph{only one} can be true except a special case: If a quantity shows both gap scaling and fluctuation scaling, then this automatically implies $\alpha = 1$. One can reverse this argument: If for a quantity one finds the relationship Eq. \eqref{eq:taylor}, with $\alpha < 1$ then it cannot exhibit gap scaling.

The proof is straightforward. Any moment of $n$ can be calculated as
\begin{eqnarray}
\ev{n^q|A} = \int_{n_0}^\infty dn n^q P(n|A) \simeq K_qA^{q\Phi},
\label{eq:nq}
\end{eqnarray}
where "$\simeq$" denotes asymptotic equality and $K_q > 0$. From Eq. \eqref{eq:nq} it follows that
\begin{eqnarray}
	\ev{\sigma^2|A} = \ev{n^2|A}-\ev{n|A}^2 \simeq \nonumber \\ K_2 A^{2\Phi}-K_1^2 A^{2\Phi}=(K_2-K_1^2)A^{2\Phi}.
\label{eq:sigma}
\end{eqnarray}
We combine Eqs. \eqref{eq:taylor}-\eqref{eq:sigma}, eliminate $A$ and find that now $\ev{\sigma^2|A}\propto\ev{n|A}^2$, i.e., $\alpha = 1$. 

The only possibility for the coexistence of gap scaling \eqref{eq:gapscaling} and fluctuation scaling \eqref{eq:taylor} with $\alpha < 1$ is when the constant factor in the variance vanishes: $$(K_2-K_1^2)=0.$$ In this case the gap scaling form does not describe the variance, that is instead given by the next order (correction) terms. Nevertheless, even if it is so, the \emph{leading} order of the variance is still zero, and consequently $F$ is proportional to a Dirac-delta: $$F\left(\frac{n}{A^{\Phi}}\right) \propto \delta\left(n/A^{\Phi}-K_1\right).$$ This case is pathological, and it is usually not considered as scaling. The conclusion: If a quantity shows gap scaling with a scaling function which is not fully degenerate (not a Dirac-delta), it must follow $\alpha = 1$.

For example, in ecology there do exist species with $\alpha \approx 1$, for which a gap scaling form of the probability density of $n$ could be valid. However, this value is by no means universal (cf. Fig. \ref{fig:fs} and Ref. \cite{taylor}). Similarly, $\alpha < 1$ was observed for Internet router traffic \cite{barabasi.fluct} or the traded value on stock markets \cite{eisler.non-universality}. These quantities cannot have a gap scaling form. Along the same lines it is possible to show that even the assumption of multiscaling leads to $\alpha = 1$, and so multiscaling can also be ruled out in cases when $\alpha < 1$.

An earlier version of the manuscript contained criticism of Ref. \cite{banavar.ecology}. We are grateful for a correspondence with Jayanth R. Banavar and Andrea Rinaldo, which has clarified why their study \cite{banavar.ecology} does not contradict the above considerations. The authors thank Marm Kilpatrick for providing ecological data. Support by OTKA K60456 is acknowledged. 

\bibliographystyle{unsrt}
\bibliography{taylorism}

\begin{thebibliography}{10}

\bibitem{stanley.phase}
H.E. Stanley.
\newblock {\em Introduction to Phase Transitions and Critical Phenomena}.
\newblock Clarendon Press, Oxford, 1971.

\bibitem{banavar.ecology}
Jayanth~R. Banavar, John Damuth, Amos Maritan, and Andrea Rinaldo.
\newblock Scaling in ecosystems and the linkage of macroecological laws.
\newblock {\em Phys. Rev. Lett.}, 98:068104, 2007.

\bibitem{bak.soc}
Per Bak, Chao Tang, and Kurt Wiesenfeld.
\newblock Self-organized criticality: an explanation of 1/f noise.
\newblock {\em Phys. Rev. Lett.}, 59:381--384, 1987.

\bibitem{taylor}
L.R. Taylor.
\newblock Aggregation, variance and the mean.
\newblock {\em Nature}, 189:732--735, 1961.

\bibitem{taylor.insect}
L.R. Taylor.
\newblock Synoptic dynamics, migration and the rothamsted insect survey.
\newblock {\em Journal of Animal Ecology}, 55:1--38, 1986.

\bibitem{kilpatrick.ives}
A.~M. Kilpatrick and A.~R. Ives.
\newblock Species interactions can explain \protect{Taylor}'s power law for
  ecological time series.
\newblock {\em Nature}, 422:65--68, 2003.

\bibitem{anderson.variability}
R.M. Anderson, D.M. Gordon, M.~J. Crawley, and M.~P. Hassell.
\newblock Variability in the abundance of animal and plant species.
\newblock {\em Nature}, 296:245--248, 1982.

\bibitem{barabasi.fluct}
M.A. de~Menezes and A.-L. Barab\'asi.
\newblock Fluctuations in network dynamics.
\newblock {\em Phys.\ Rev.\ Lett.}, 92:28701, 2004.

\bibitem{duch.internet}
Jordi Duch and Alex Arenas.
\newblock Scaling of fluctuations in traffic on complex networks.
\newblock {\em Phys. Rev. Lett.}, 96:218702, 2006.

\bibitem{eisler.non-universality}
Z.~Eisler, J.~Kert\'esz, S.-H. Yook, and A.-L. Barab\'asi.
\newblock Multiscaling and non-universality in fluctuations of driven complex
  systems.
\newblock {\em Europhys. Lett.}, 69:664--670, 2005.

\end{thebibliography}

\end{document}